# Straw-based coordinate muon chamber

V.D. Peshekhonov[a,*], K.I. Davkov[a], V.V. Myalkovskiy[a], N.A. Russakovich[a], I.A. Zhukov[a]

[a] *Joint Institute for Nuclear Research (JINR), Dubna, Russia*

**Abstract**

The article considers a prototype module with sensitive dimensions of 2 x 0.5 m$^2$ based on 2 m long straw tubes which preserves its geometrical dimensions up to the pressure of 4 bar independently of the ambient temperature and humidity. The suggested technique provides a possibility of constructing relatively low-cost planar modules by using straw tubes as long as needed for constructing large area detectors at colliders or other field experiments by assembling them in a common coordinate detector with a user-specified acceptance. The distinguishing feature of a detector is its good radiation hardness, low radiation thickness and a possibility of optimizing the detector operation mode in a large range of gas pressure.

*Corresponding author
*E-mail*: pvd@sunse.jinr.ru

## 1. Introduction

Gaseous detectors are widely employed in experiments at particle accelerators. They include drift chambers (DC) with cathode readout, resistive plate chambers (RPC), thin gap chambers (TGC) and coordinate detectors based on aluminum drift tubes (Al-DT) of large diameter which are being used as components of muon detectors at the CERN LHC [1--3]. Al-DT can operate in a broad range of the filling gas pressure but they are characterized by large collection time of the ionization electrons and relatively low counting rate due to distortion of the electric field caused by volume charge of slowly drifting positive ions. Aluminum tubes make it possible to construct large area detectors which arouses interest in employing them in high-angular resolution muon detectors for radiographic and tomography studies of large-scale objects [4,5]. However, the possibility of improving parameters of the DT-based coordinate detectors by means of decreasing the tube diameter is limited by the technological difficulty in producing long tubes of high straightness and uniformity of their diameter.

The coordinate detectors based on thin-film drift tubes (straws) of diameter from 4 to 15 mm are well known. The straws have a good straightness and highly uniform inner and outer diameters independent of their length [6,7]. Straws allow one to reach larger granularity by decreasing their diameter and owing to a possibility of using segmented anodes [8,9]. Moreover, they have a good spatial resolution [10,11] and low radiation thickness. The ratio of a radiation thickness of an Al tube with a wall of 0.4 mm to that of a typical straw tube is more than 18 which indicates to a possibility of increasing constructive radiation thickness of straw-based detectors.

This article considers a one-layer prototype module based on 2 m long straws with the carbon cathode. Sensitive area of the prototype is 2 x 0.5 $m^2$, and its size remains unchanged up to 4 bar of the operating gas mixture pressure and not affected by variations of environmental parameters such as humidity and temperature.

## 2. Design

The prototype has been designed by considering a feasibility of constructing standard modules convenient for assembling them into large area detecting systems that can operate at high pressure of filling gas.

The developed one-layer prototype has a rigid planar structure containing N (multiple of 8) straws covered with epoxy resin. The structure thickness is only 0.2 mm larger than the straw diameter due to the increase of 0.1 mm on each surface. The production technique makes it possible to preserve both the straightness and straw diameter as well as a constant gap between adjacent straws. This has been achieved by partially employing the technique developed earlier at JINR (Dubna) for construction of the straw tracking system for the COMPASS detector [11,12]. The prototype contains 48 straws with an inner diameter of 9.56 mm, 2 m length and a wall thickness of ~60 μm. An aluminum thin-wall profile is glued in parallel direction with the straw orientation to each side of the straw tube module. The profile height is larger than the structure thickness by the value of $h$ where $h$ is the diameter of two metallic pipes used as gas manifolds (GM).

After production of the straw planar structure the insulating end-plugs with the ground springs and the crimping pins as well as the anodes are inserted into the straws. The tension of the anode made of 30 μm gold-plated tungsten wire is 70 g, and each anode has one central spacer. Each straw has been tested by using a $Fe^{55}$ source under flush of an operating gas mixture separately from other straws. Afterwards, two tubular GM are mounted on one side of the structure near straw ends. The manifolds are used for the supply and return of the gas mixture, and they are joined together with side profiles in a common frame and are covered with epoxy resin later on. The diagram displaying an installation of the GMs is shown in Fig. 1. The manifold is connected with an inner volume of each straw by using metallic capillary tubes passing through holes made beforehand in the end plugs and the straw walls and are placed orthogonally to the straw plane.

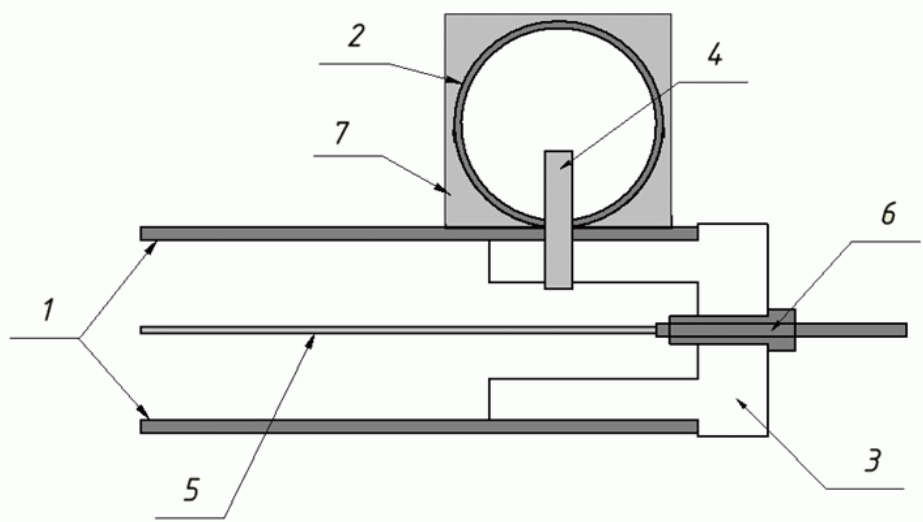

Fig. 1. Schematic layout of the gas manifold mounted on straw planar structure. 1 - straw walls, 2 - Al tube of 8 mm diameter and 0.4 mm wall thickness, 3 - plastic end-plug, 4 - metal capillary tube, 5 - anode, 6 - crimping pin, 7 - epoxy resin.

This surface is laminated with the metalized film which serves both as the electromagnetic screen and protection against high humidity. The supporting carbon or fiberglass strips of height $h$ shown in Fig. 2 are mounted on the module surface with the interval determined by the straw length. Three fiberglass strips of 0.2 mm thickness and 8 mm height separated by equal distance are mounted on each side of the prototype module which is not supposed to be assembled into a two-layer chamber. The motherboards (MB) and termination boards (TB) are mounted near crimping pins, and the anodes are galvanically connected with corresponding buses on the boards. The end parts of the straws and end plugs are hermetically sealed with an epoxy resin. Therefore, the prototype module has two tubular GMs on one side which serve as the parts of its frame. In the adopted design scheme, the GMs do not contain any material which would contaminate gas mixture, and the MBs and TBs are connected to anodes without passing through hermetic gaseous volumes.

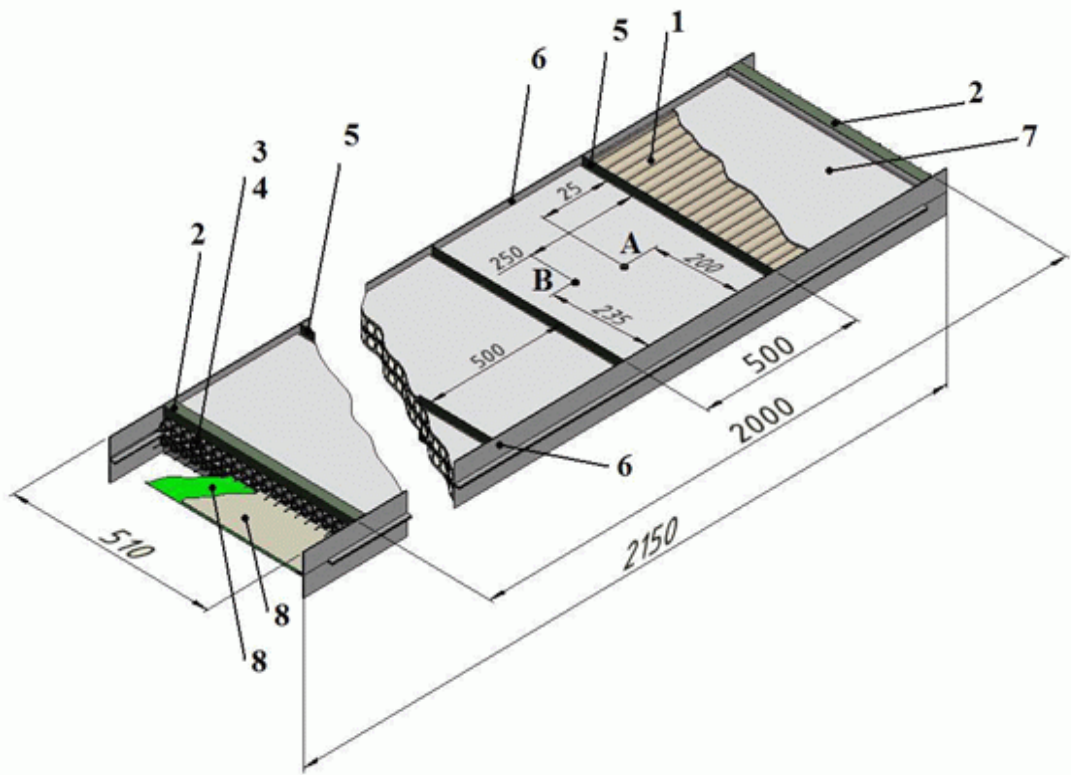

Fig. 2. Layout of the two-layer detector. 1 - straw, 2 - gas manifolds covered by epoxy resin, 3, 4 - end-plugs with crimping pins, 5 - supporting strips, 6 - thin-walled profiles, 7 - metallized film, 8 - mother boards.

The module side opposite to GMs serves for gluing on two identical modules in one mechanical unit with distributing tubular GMs, MBs and TBs. Upon gluing, the modules are shifted relative each other by one straw radius. The common frame of such two-layer chamber may contain fixing holes to combine several chambers in one detecting system. The general layout and some geometrical dimensions of the double-layer chamber involving two identical modules is shown in Fig. 2.

## 3. Mechanical parameters of the module.

The module surface swelling in the orthogonal direction as a function of the gas pressure has been measures in points A and B shown in Fig. 2. Point A is located in the center of the region bound by two strips, and point B is near a strip. Upon increasing gas pressure, the rapid increase of the straw dimension in the direction orthogonal to the module has been observed (see Fig. 3). Maximal increase in the module thickness in point A is 260 μm (per straw radius) at the pressure 4 bar. During about 2.5 hours one observes a slow increase in the thickness up to about 350 μm which is ~7% of the straw radius. The thickness decreases rapidly to 100 μm above the nominal value after pressure release and returns to normal one after about 2.5 hours. The corresponding deformation in point B is found to be smaller by factor 1.6 and 1.8 for the pressure 4 and 3 bar, respectively. These changes result in a slight deterioration of the straw circular parallelism which is typically smaller than the maximal admissible non concentricity of the anode-cathode position (10% for the TRT ATLAS straws [13]). The magnitude of the expansion can be reduced by decreasing the interval between supporting strips or by an increase of the module thickness.

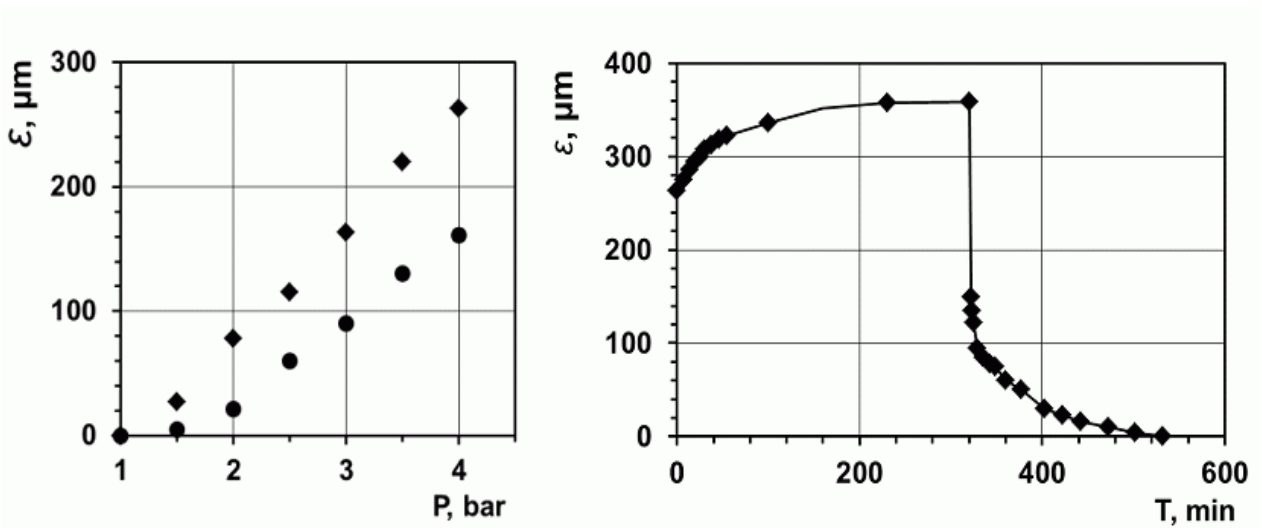

Fig. 3. Displacement of module surface (ε) due change of pressure of gas filling from atmospheric to 4 bar in point A (■) and point B (●).

Gas tightness check for the prototype has demonstrated that the gas leak at 4 bar was about 2.4 mbar/h which has been caused by a leaky joint between the Cu-tube and plastic covering of the crimping pins. Epoxy sealing of these seams removes the leak.

The radiation thickness of the straw is less than 0.05% $X_0$. It is instructive to compare this parameter for the straw based detector and the detector based on aluminum tubes. Assuming the same radiation thickness for the two types of detectors, the relation between the wall thickness h for the aluminum tubes and the straw diameter d is displayed in Fig. 4. It is safe to say that the increase of radiation thickness of the constructed module with the straws of small diameter (4-6 mm) is negligible.

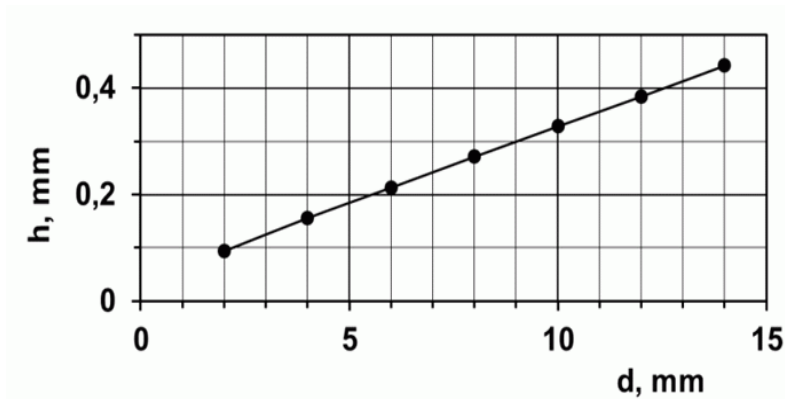

Fig. 4. The relationship between Al tube wall thickness h, and straw diameter d, for the same radiation thickness of a high-pressure gaseous detector.

**4. Prototype parameters in the pressure operation region**

The gas mixture $ArCO_2$ (80/20) and $Fe^{55}$ irradiation source have been used in the tests of the prototype module with the pressure of the gas varied from 1 to 3 bar at both the normal and special operating mode where the spatial resolution better 50 μm with the efficiency 99% was earlier measured [14].

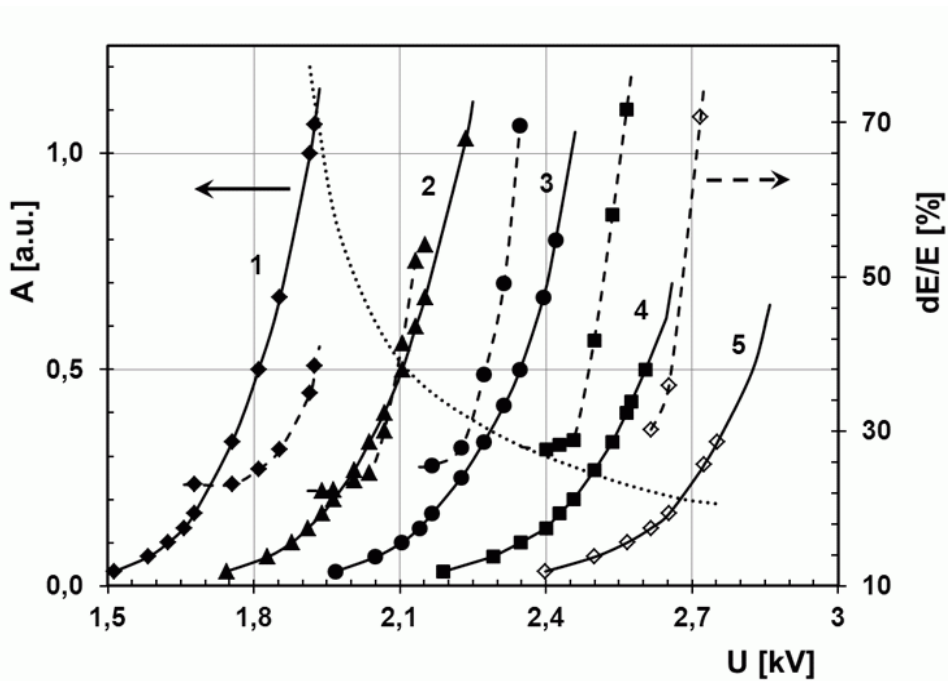

Fig. 5. Signal amplitudes in relative units produced by 5.9 keV gammas (solid line) and the energy resolution (dashed line) as a function of anode voltage. The unity in the ordinate axes corresponds to gas amplification ~$6\times10^4$. The pairs of curves with numbers from 1 to 5 correspond to gas mixture pressure of 1, 1.5, 2, 2.5 and 3 bar, respectively. The energy resolution is better than 40% for the signal amplitudes which are below the boundary shown with dot-dashed line.

The signal amplitude and energy resolution of the straw as a function of the anode voltage at a gas pressure in the range from 1 to 3 bar are shown in Fig. 5. One can see a rapid deterioration of the energy resolution with gas pressure rising. The signal amplitudes corresponding to the 40% energy resolution are shown with dot-dashed line. For this energy resolution and the design option involving straws with 30 μm anode wires and the gas mixture in question, the signal amplitudes decrease by factor larger than 2 and 5 if the pressure rises from 1 to 1.5 and 3 bar, respectively. The collection time of the ionization electrons varies from ~100 to ~200 ns with the pressure changing from 1 to 4 bar.

The spectra of the signals from the straw at 1 bar pressure and anode voltage 1.79 and 1.87 kV are shown in Fig. 6. One can see a good separation of the main pulse of energy release and the escape peak for the gas mixture pressure of 1 bar. The energy resolution deteriorates with increasing high voltage, moreover, afterpulse can be generated that causes triggering of a registration circuit in the case of high gas amplification (see right panel in Fig. 6 for gas amplification ~$6 \times 10^4$).

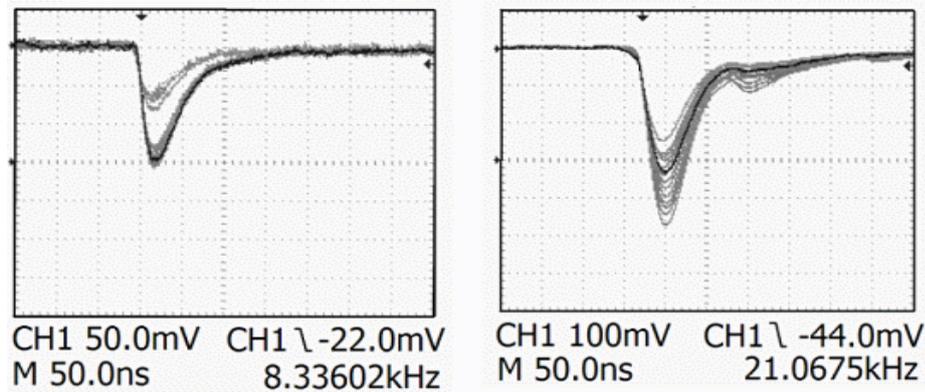

Fig. 6. Signal amplitudes from the straw under the anode voltage 1790 V (left panel) and 1970 V (right panel) which features afterpulses arising with the increase of the voltage. The gas mixture is under pressure of 1 bar.

The working range of the straw in the proportional mode at high pressure of the gas mixture decreases as well. The fragments of signal spectra generated by 5.9 keV gammas for gas filling pressure 3 bar are shown in Fig. 7. The ratio of maximal to minimal signal amplitude is about 3 for high voltage 2.75 kV, and the signal in full range of amplitudes are observed upon further increase of the voltage. In addition, a high-current signals arise and reach 20% for the voltage ~2.85 kV. The number of such signals increases to about 45, 70 and 75% upon increasing the voltage to 2.95, 3.05 and 3.1 kV, respectively. The large dynamical range of indicates to transition mode of the detector operation from the proportional and space charge saturation regime to the self-quenching streamer (SQS) mode.

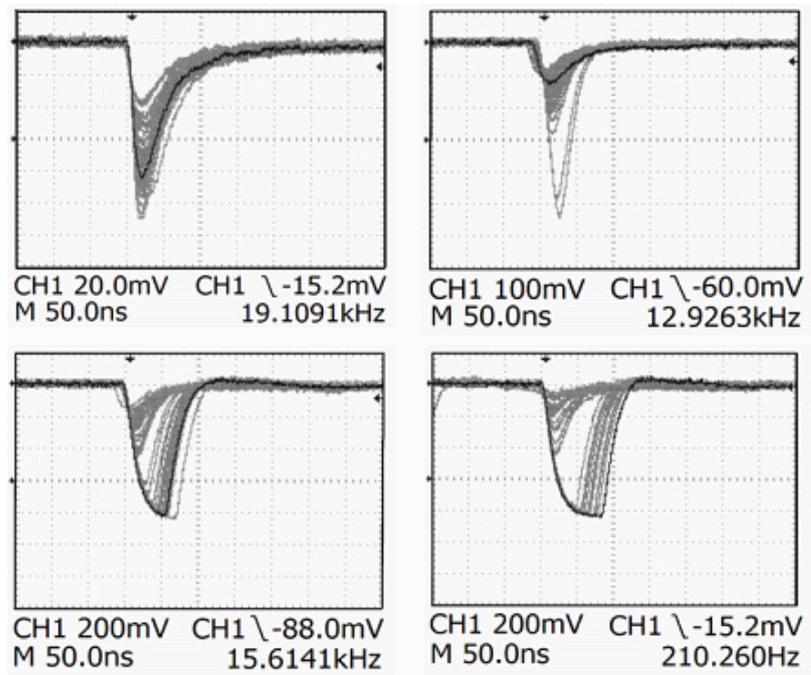

Fig. 7. Signal amplitudes from the straw at the pressure 3 bar for the anode voltage 2.75 kV (top left panel), 2.85 kV (top right panel), 3.05 kV (bottom left panel), and 3.1 kV (bottom right panel).

The measurements of the signals induced by the irradiated straw on the adjacent channels have demonstrated that the cross-talk is about 0.6%. Figure 8 shows typical signals from the irradiated (upper signal) and adjacent (lower signal) straw for the pressure of 3 bar at the anode voltage 2.9 kV by displaying a pulse of normal amplitude (left panel) and high-current event (right panel). One can see that the ratio of the unipolar signals is also below 0.6%.

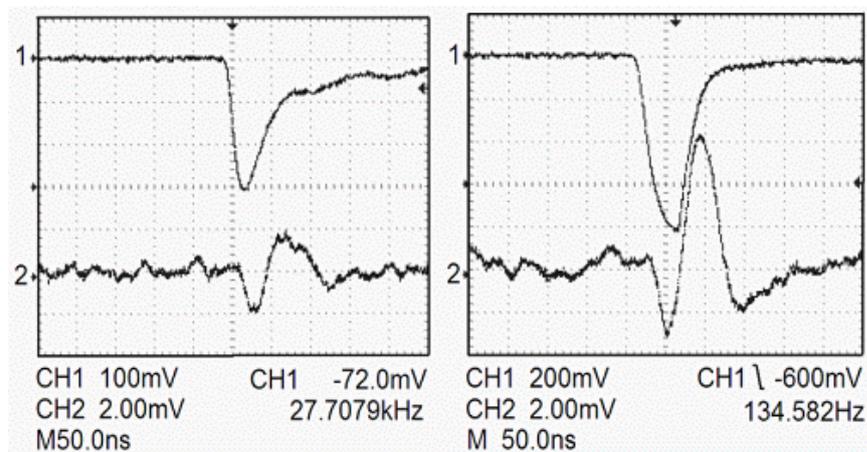

Fig. 8. The straw cross-talk signals for the anode voltage of 2.9 kV. The pulse of normal amplitude is displayed in the left panel, the high-current event is shown in the right panel. The upper curve corresponds to a signal from the irradiated straw, the lower curve shows a signal read out from a neighboring straw.

Counting-rate curves for the gas filling at 1 and 3 bar and the registration thresholds about 3, 4.5 and 8 fC are displayed in Fig. 9. Upon increasing gas filling pressure to 3 bar, the efficiency of gamma detection is increased by a factor of $k_p \approx 2.4$. A weak dependence of the noise level beginning from the discriminator threshold to the anode voltage 1.95 is found at the pressure of 1 bar. The detector can work at the pressure 3 bar in the saturated proportional regime and in the beginning of the SQS mode to produce the high-current signals. Thus, the low-threshold registration of events is feasible if the dark current noise is low and cross-talk is not high. For high discriminator thresholds of ~8 and ~12 fC the efficiency loss of about 3 and 9% is observed which indicates the presence of low-amplitude signals.

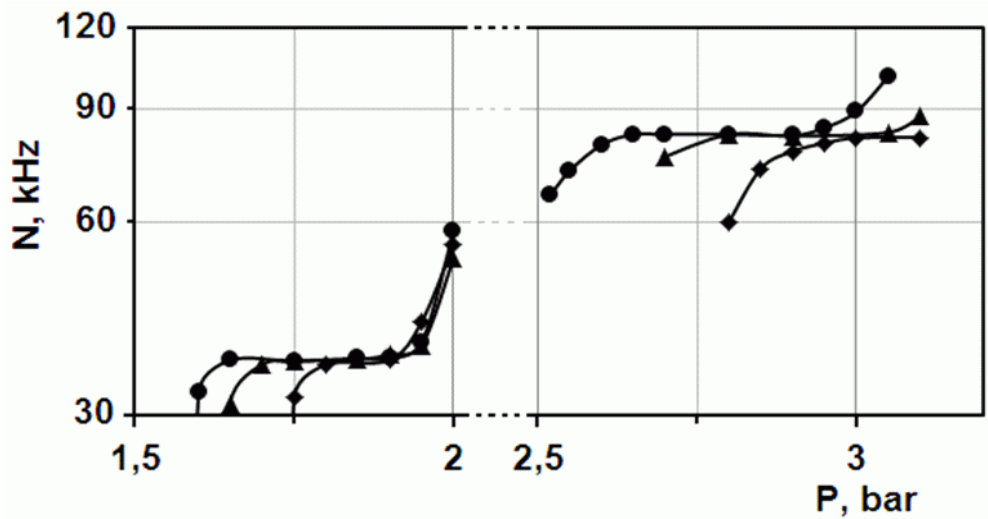

Fig. 9. Counting rates measured for three values of the discrimination threshold and for gas filling pressure 1 and 3 bar. The curves on the left-hand side correspond to the pressure of 1 bar, and those on the right-hand side correspond to the pressure 3 bar. The data obtained for three different discrimination thresholds ~3 fC, ~4.5 fC and ~8.5 fC are displayed with diamonds, circles and triangles, respectively.

The drift tubes are characterized by fairly good radiation hardness [7,15--17] in the proportional and limited-proportional modes of operation. The straws identical to those used in the prototype under pressure 3 bar and 3.05 anode voltage have demonstrated high spatial resolution in registration of minimum ionizing particles (MIP) [14]. The feasibility of long term operation of the straws in this regime is verified in aging test. The preliminary test involved an X-ray tube at the voltage of 9 kV ($I_{Rö}=3\mu A$) which irradiated a 11 cm long straw, and irradiation of the straw was near 2.5 C/cm. Subsequent tests by careful scan of the straws with the collimated $Fe^{55}$ source have demonstrated that both the signal amplitudes and energy resolution had not changed in the irradiated region. These results indicate the possibility of using detectors constructed on the basis of the straws under discussion in the environment characterized by high rates of charged particles with different operating mode.

## 5. Conclusion

The technique developed for the prototype makes it possible to construct relatively low-cost planar modules by using straws of large length as well as to assemble them in a common coordinate detector with a user-specified acceptance. If necessary, the straws of diameter 4 mm and larger can be granulated [9]. A good radiation hardness, low radiation thickness and a possibility of operation with the gas filling from normal pressure up to 4 bar by keeping the geometrical dimensions unchanged offer additional opportunities for optimization the mode of detector operation including feasibility to work in proportional or limited proportional modes. In particular, it is of interest to improve the spatial resolution by employing coordinate detectors of the MIP in the high current operating mode.

**References**

[1]. G. Aad et al., JINST 3:S08003, 2008.

[2]. C. Adorisio et al., Nucl. Instr. and Meth. A 598 (2009) 400-415.

[3]. Albert Engl et al., arXiv:0908. 2507.

[4]. V. Anghel et al., Nucl. Sci. Symp. Conf. Record (NSS/MIC), IEEE, 6 (2010) p. 547.

[5]. C. L. Morris et al., Science and Global Security, 16:37–53, 2008.

[6]. P. Abbon et al., Nucl. Instr. and Meth. A 577 (2007) 455.

[7]. E. Abat et al., JINST 3: P10003, 2008.

[8]. K. Davkov et al., Nucl. Instr. Meth. A 584 (2008) 285.

[9]. S. N. Bazylev et al., Nucl. Instr. and Meth. A 632 (2011) 75.

[10]. T. Akesson et al., Nucl. Instr. and Meth. A 485 (2002) 298.

[11]. V. N. Bychkov et al., Part. and Nucl. Lett., 2002, №2 | 111 |, p.p.64-73.

[12]. V. N. Bychkov et al., Nucl. Instr. and Meth. A 556 (2006) 66-79.

[13]. Yu. V. Gusakov et al., Phys. Part. Nucl. 41:1-26, 2010.

[14]. V. I. Davkov et al., Nucl. Instr. Meth. A 634 (2011) 5-7.

[15]. T. Akesson et al., Nucl. Instr. and Meth. A 515 (2003) 166.

[16]. M. Deile et al., Nucl. Instr. and Meth. A 518 (2004) 65.

[17]. B. Bittner et al., Nucl. Phys. B (Proc. Suppl.) 215 (2011) 143.